# Glassiness in vortex matter in $MgB_2$ probed by a novel scaling method for creep analysis.

G. Pasquini[*] and A. Moreno

*Departamento de Física, FCEyN, Universidad de Buenos Aires; CONICET; Pabellon 1, Ciudad Universitaria, Buenos Aires, Argentina.*

A. Serquis and G. Serrano

*Instituto Balseiro- Centro Atómico Bariloche, CONICET, S. C. de Bariloche, 8400, Río Negro, Argentina.*

L. Civale

*Superconductivity Technology Center, Los Alamos National Laboratory, MS K763, Los Alamos, New Mexico 87545, USA.*

(Dated: September 3, 2009)

## Abstract

The glassy dynamics is a consequence of the elastic properties of the vortex matter, and in principle may occur in any superconductor. However, whereas a large amount of experimental evidence confirms glassiness in high Tc superconductors (HTS), the applicability of the whole framework developed to describe vortex matter in HTS to other superconducting materials is unclear. In this framework, the particular creep behavior of $MgB_2$, larger than creep of conventional superconductor materials but much lower than HTS one, had precluded a complete understanding of the vortex regimes in this material. In this work we present an experimental study of relaxation processes of pure $MgB_2$ bulk samples measured by DC magnetization technique. We propose a novel self-consistent scaling method to analyze the data and extract the activation energies. The observed experimental behavior can be described in a glassy picture, with a unique critical exponent $\mu \sim 1$, characteristic of collective creep in the intermediate vortex bundle regime.

## I. INTRODUCTION

Since the discovery of the oxide high temperature superconductors (HTS) vortex physics became a major field in condensed matter and statistical physics. Besides the technological relevance, the continuous interest in the topic has been motivated by a novel and fascinating phenomenology. Vortices in the HTS exhibit a complex equilibrium phase diagram, with a proliferation of crystalline solid, glassy and liquid phases separated by thermal and dimensional first and second order phase transitions, as well as a rich dynamic behavior in non-equilibrium conditions, including very fast and glassy flux creep. A fact that is not always appreciated, however, is that there is no hard boundary between the properties of vortex matter in cuprates and other superconductors. Certainly thermal fluctuations play a far more important role in HTS than in low temperature materials due to the combination of the small coherence length (which is a direct consequence of the high critical temperature $T_c$) and the large electronic mass anisotropy. This determines that vortex liquid phases occupy large portions of the temperature - magnetic field plane and also promotes the rapid relaxation of metastable states, but is not a requirement for glassiness.

The decay of non-equilibrium current density due to thermal fluctuations, known as flux creep phenomenon, refers to thermally activated motion of vortices at current density $J$ lower than the critical current density $J_c$. From a practical point of view, the optimization of tapes and wires in new superconducting materials requires a deep understanding of flux creep. Therefore, a large amount of studies of the time decay of the current as a function of temperature $(T)$ and magnetic field $(B)$ have been performed performed to understand the vortex dynamic. In all the cases, the important quantity to know is the activation energy barrier $U(J; T; B)$.

In traditional type II superconductors the creep phenomenology is well known since nearly 50 years ago: in this case the current decay is very small, so $(J_c - J) << J_c$, and this leads to the traditional Anderson-Kim (A-K) model[1]. The discovery of high temperature superconductors (HTSs) and their "giant" flux creep opened a new chapter in vortex dynamics. A large number of glassy creep regimes in the limit $J << J_c$[2,3] have been proposed theoretically and many of them observed experimentally[4]. The glassy dynamics is a consequence of the elastic properties of the vortex matter, which under certain conditions results in activation energies for depinning that diverge in the limit of vanishing current density, and in principle may occur in any superconductor. However experimental confirmation of glassiness, which regardless of the specific approach always involves detecting tiny deviations from the non-glassy flux creep described by the Anderson-Kim model, is extremely challenging in strong pinning low $T_c$ superconductors, due to the very slow creep. As a consequence, the applicability of the whole framework developed to describe vortex matter in HTS to other superconducting materials is unclear.

In this context, $MgB_2$ is a very particular system. Since its discovery, a lot of work was devoted to understanding and improving[10–13] its pinning properties. Moreover, it is an ideal model system to explore vortex dynamics in "conventional" superconductors, as the intermediate $T_c$ and moderate anisotropy makes the creep effects smaller than in HTS but still large enough to allow detailed exploration. The influence of thermal fluctuations in the vortex physics is measured by the Ginzburg number $G_i = \frac{1}{2}(kT_c/H_c^2\xi^3\gamma^{-1})^2$, where $H_c$ is the thermodynamic critical field, $\xi$ is the coherence length, $\gamma$ is the pinning parameter and $k$ the Boltzman constant. For $YBa_2Cu_3O_7$ this is as large as $G_i \sim 10^{-2}$, and even larger for the more anisotropic $Bi$-based compounds, while for $NbTi$, the paradigmatic strong-pinning conventional superconductor, $G_i \sim 10^{-8}$. For $MgB_2$, depending on the doping level (that modify $\xi$ and $\gamma$, and to a lower extent $T_c$) we have $G_i \sim 10^{-4} - 10^{-5}$. This is just in the middle between the extreme cases, and in addition the easy factor-of-10 tunability in $G_i$ (e.g., by carbon doping) allows systematic manipulation of the creep effects. The first consequence is that in magnetization measurements, the measured $J$ is expected to be in the intermediate range $(J_c - J) \lesssim J_c$. As a consequence of this particular "intermediate" creep, the experimental study of relaxation processes and the corresponding activation energies in $MgB_2$ has not been completely understood up to now.

In this paper we show that flux creep in strong pinning bulk $MgB_2$ is glassy with a glassy exponent $\mu$ that is consistent with creep by vortex bundles, as frequently observed in HTS. We achieved this by developing a novel self-consistent scaling method. The analysis is based in the well known Maley analysis[5] of the activation energy as a function of current density but takes into account the intermediate $J$ range and incorporates the effects of both $J_c$ and $U_c$ temperature dependences.

The paper is organized as follows: In Section II, a brief review of the main concepts related to activation energies and Maley analysis is done. In Section III, experimental details are given. In Section IV, typical relaxation data, together with the scaling method are presented. Results are discussed in Section V. Finally the conclusions are sketched in section VI

## II.  CREEP AND ACTIVATION ENERGIES

In traditional type II superconductors the current decay is very small, so $(J_c - J) << J_c$, therefore

$$U(J,T,B) \simeq U_c(T,B)(1 - \frac{J}{J_c(T,B)}) \qquad (1)$$

This leads to the traditional logarithmic time decay of current (i.e. magnetization) of Anderson-Kim (A-K) model. The pinning energy $U_c(T,B)$ may be experimentally extracted from the relaxation rate $S = -d(\ln(J))/d(\ln(t)) \simeq kT/U_c$.

On the other hand, in HTS´s, a variety of glassy creep regimes in the limit $J << J_c$[2,3] have been theoreticaly proposed. In all the cases

$$U(J,T,B) \simeq U_c(T,B)(\frac{J_c(T,B)}{J})^\mu = g(T,B)J^{-\mu} \qquad (2)$$

where $\mu$ is a critical exponent that depends on the particular regime. A mayor difficulty to directly extract the parameters $U_c$, $J_c$ and particularly $\mu$ from the relaxation data is the extremely large time needed to reliably determine the $\mu$ exponent. To overcome this problem, Maley et al.[5] proposed a scaling method, valid in the limit $J \ll J_c$ , that collapses the relaxation curves collected at various temperatures and allows to fit a general function

$$U(J,T,B) = g(T,B)f(J) \sim Cf(J) \qquad (3)$$

in a large range of $J$. The Maley analyisis makes no assumptions a priori about the functional dependence of $U$ and $J$. Thus, the experimentally found $U(J)$ can be used to determine whether or not the relaxation is glassy (i.e. Eq. 2 is valid) and, if so, to obtain $\mu$. Whereas the original method has been proposed to analyze magnetic relaxation and it is valid in the limit where $g(T,B)$ is nearly independent on temperature, further generalized Maley procedures have included $T$ and $B$ dependencies allowing extending the method to ac susceptibility measurements[6] and to higher temperatures and fields. The validity of formula (2) in HTS has been proved in a variety of experiments.

Therefore, an interpolation formula valid in both limits, (1) and (2), has been proposed[2,8]:

$$U(J,T,B) \simeq U_c(T,B)[(\frac{J_c(T,B)}{J})^\mu - 1] \qquad (4)$$

In this general case

$$\frac{1}{S} = \frac{U_c(T;B)}{kT} + \mu\ln(1+t/t_0) \sim \frac{U_c(T;B)}{kT} + \mu\ln(t/t_0) \qquad (5)$$

The last expression has been frequently used[9], to extract the critical exponent.

As was mentioned in the introduction, as a consequence of the "intermediate" current range $(J_c - J) \lesssim J_c$, the experimental study of relaxation processes in $MgB_2$ is particularly complex. At high temperature, near $T_c$, some attempts to apply a generalized Maley Method in ac susceptibility measurements have been done[14], but the results are not coincident. Although at low temperatures the A-K model has been usually applied to estimate $U_c$ [15], the full expression 4 must be applied. Recently an attempt has been proposed by Miu et al[21]. In this analysis, they neglected the $T$ variation of $J_c$, $U_c$ and critical exponent, but they ascribed the big experimental $M(T)$ dependence to a $T$ dependent overall relaxation before the time at which the first data point is taken, hard to justify under their assumption.

The analysis proposed in the present paper is based in the Maley analysis but takes into account the full expression 4 and incorporates the effects of both $J_c$ and $U_c$ temperature dependences.

### III. EXPERIMENTAL

$MgB_2$ samples used in this study are prepared by solid-state reaction with Magnesium (-325 mesh, 99%) and amorphous boron (99%) as starting materials[18]. The powders were ground inside a glove box and pressed under $\sim 500\ MPa$ into small pellets with dimensions of 6 $mm$ in diameter and $\sim 4\ mm$ in thickness, wrapped together with extra 20%at $Mg$ turnings (99.98% Puratronic) in $Ta$ foil and then placed in an alumina crucible inside a tube furnace in flowing $Ar/H_2$ at 900∘C for 30 $min$.

The magnetization measurements were carried out in a Quantum Design model MPMS XL 7T SQUID based magnetometer. Time-dependent data were taken with a protocol similar to that described in ref.[9]. A scan length of 3 $cm$ was used in order to minimize the effects due to the nonuniformity of the applied magnetic field, that was applied parallel to the longest axis of the sample. For each relaxation measurement, the samples were first cooled and stabilized at the measurement temperature. Then the field was first raised up to $6T$ and then lowered to the measuring field to assure that the sample was fully penetrated. Intermediate measurements were performed in the upper and lower magnetization branches to substrate the reversible magnetization. We checked for and ruled out any effects due to the magnet self relaxations ($H$ variations during the measurement time) that could lead to spurious changes in the magnetization of the sample*s*.

The current density $J(t)$ has been calculated from the measured magnetic moment, using the relationship for a sample with rectangular cross section in the critical state following the Bean model[16].

## IV. DATA ANALYSIS:

In the following all the data analysis are performed at constant applied $H$; therefore, with the aim to simplify the notation, the $B$ dependences are omitted in all the formulas. Figure 1 shows the relaxation rate $S$ as a function of temperature, for data measured at $H = 1T$. In the insets, two sets of experimental relaxation curves $J(t)$, taken at several temperatures $T$, spaced in $\Delta T = 1K$ at low (inset a) and high (inset b) temperature are shown in double logarithmic scale. The relaxation is clearly observable but there is a big jump $\Delta J$ between adjacent temperatures. Any attempt to perform a Maley analysis using physically reasonable $C$ and $g(T)$ fails. The key point to explain this fact is that, in this system, even at low temperature, the $J_c(T)$ dependence in (4) cannot be disregarded. In the present case, the main cause of the temperature dependence in the measured current $J(T)$ dependence is not the different time windows during measurements, but the intrinsic $J_c(T)$ dependence.

To analyze the data, we have developed the following method:

As in the Maley method, we use the general creep relationship

$$\frac{U(J,T)}{kT} = -\ln\left|\frac{dJ}{dt}\right| + C$$

However, in our case, we **assume** the validity of the full dependence (4)

$$\frac{U_c(T)}{k}[(\frac{J_c(T)}{J})^\mu - 1] = -T\left[\ln\left|\frac{dJ}{dt}\right| + C\right]$$

so

$$J^{-\mu} = \left[-T\ln\left|\frac{dJ}{dt}\right| + CT + \frac{U_c(T)}{k}\right]\frac{kJ_c(T)^{-\mu}}{U_c(T)} \quad (6)$$

Therefore, at each temperature

$$-\mu\ln J \propto \ln\left[-T\ln\left|\frac{dJ}{dt}\right| + E(T)\right] \quad (7)$$

where $E(T) = CT + U_c(T)/k$.

We now proceed as follows: for each $T$, we pick a number $E(T)$. We then plot $(-T\ln\left|\frac{dJ}{dt}\right| + E(T))$ vs $J$ in double log scale. The data for each $T$ will form a short segment, similar to the case in the traditional Maley method. If the glassy description (7) holds with a unique $\mu$ in all the $T$ range analized, it will be possible to find a suitable set of numbers $E(T)$ such that the segments for all $T$ have the same slope $-\mu$.

The uniqueness of the solution is not obvious. Therefore, some reasonable physical conditions may be added to find the good choice. The typical assumption of a $T$ independent $U_c$ at low temperature is not necessarily valid, because in this system there is an anomalous non linear $S(T)$ in this range (see figure 1). A more general condition is given by searching the best self-consistence with Eq. (5). The exponent $\mu$ and energies $U_c(T)/k$ are self-consistently found in a way to satisfy both (5) and (7). From values found in the literature[5-7,9], we fixed the parameters $\ln(t/t_0) = 29$ in (5) and $C \backsim 24$ in (7)[7].

For all relaxation data taken at $H = 1T$ between $5K$ and $25K$, the best self-consistence is achieved with an exponent $\mu = 1.00 \pm 0.03$. In the insets of Figure 2, examples of the good fit achieved at low (inset a) and high temperature (inset b) are shown. Once $\mu$ and $E(T)$ have been obtained, in order to collapse all the data in an unique linear function, a constant term $D(T)$ has been added at each temperature. The main panel in Figure 2 shows the scaling linear function $f = \ln\left[-T\ln\left|\frac{dJ}{dt}\right| + E(T)\right] + D(T)$ obtained. The linearity is a clear advantage, because the scaling of non-linear data without overlapping between them (as is the present case) could be doubtful.

The consistence of critical exponent and energies is shown in Figure 3a. Black squares indicate the values of $U_c(T)/k$ obtained from the parameters $E(T)$ used at each temperature in order to reach the linear relationship (7) with fixed sloop $\mu = -1$ for all $T$. The values obtained from the experimental $S(T)$ (Figure 1) using formula (5) (with $\mu = 1$) are compared. Data are consistent with the model above $T = 7K$. Below this temperature other factors (perhaps macroscopic jumps) increase the relaxation rate and the model breaks down. Obviously, the estimated $U_c$ values are not reliable at too high temperatures, once $U_c(T)/k \lesssim CT$.

The same analysis has been performed in relaxation data obtained at $H = 3T$. A good consistence with the same critical exponent $\mu = 1.00 \pm 0.03$, was obtained above $T = 5K$ (Figure 3b). The constancy of $\mu$ indicates that the same creep regime occurs over a wide region of the $T - H$ space, and the field range strongly suggests a collective creep scenario.

## V. DISCUSSION

Once the self-consistency is achieved, other physical parameters related to vortex dynamics may be obtained. From (6) and (7), it can be seen that the constant term added at each $T$ in order to collapse all the data in Figure 2 is $D(T) = \ln(\frac{kJ_c(T)^{-\mu}}{U_c(T)})$. Therefore, the "true" $J_c(T)$ can be extracted. Figure 4 compares the measured current density $J$ at the beginning of the relaxation (generally called the "measurable critical current") and the "true" $J_c$, at $H = 1T$ and $3T$. It can be seen that the supposed condition of an intermediate creep, with $(J_c - J) \lesssim J_c$ holds in all the range.

Observing the resulting $T$ dependence in Figures 3 and 4, it seems that pinning energies remain nearly constant at low temperatures and drop at a field dependent temperature $T_1$, far below the irreversibility line. On the other hand, a continuous decrease occurs in the critical current density. From the comparison of data taken at $H = 1T$ and $3T$, it seems that both $J_c$ and $U_c$ decrease with $B$. This field dependence for $U_c$ is not that predicted by the classical collective pinning theory for the activation energy of vortex bundles when pinning arises from random point defects[2]. However this discrepancy is not surprising, as we know that the strong pinning in these $MgB_2$ samples arises from a variety of larger defects rather than atomic-size disorder. For instance, some creep regimes associated with aligned columnar defects are also consistent with $\mu \sim 1$ and a $U_c$ that decreases with $B$[2]. The values of the $\mu$ exponents for randomly distributed nano-sized defects have not been investigated in detail, but recent results in $BaZrO_3$-doped $YBa_2Cu_3O_7$ films suggests that they may also be in the range of $\mu$ ~1[17].

The pinning volume is determined by the competition between elastic and pinning energies and each vortex bundle of volume $V_c$ is collectively pinned with an energy $U_c$. The Lorentz force over a volume $V_c$ is $F_L \backsim \frac{1}{c}BJV_c$ and, in a rough estimation, the pinning force is $F_p \backsim U_c/\xi$, where $\xi$ is the coherence length. When $J$ reaches the critical current density $J_c$, the pinning and Lorentz forces are balanced and therefore the following relationship is obtained:

$$V_c(T,B) \backsim \frac{c\ U_c(T,B)}{B\ \xi(T)J_c(T,B)}$$

Notice that this is a very general expression. From $H_{c2}$ measurements, we have recently shown[18] that $\xi(T)$ fits very well the function proposed in ref.[19,20], with a $\xi(0) \sim 50A$. Using this function for $\xi(T)$ we have estimated $V_c(T,B)$ from our data. Results are shown in Figure 5, where $V_c$ is plotted as a function of temperature for $H = 1T$ and $3T$.

The estimated numerical values ($\sim 10^{-15} cm^3$) are in agreement with the supposition of vortex bundles, because the correlation radius is larger than the main vortex distance. Moreover, $V_c$ is on the order of $\lambda^3$, in coincidence with the regime of intermediate bundles.

The sudden drop of $U_c(T)$ at $T_1$ is caused by a decrease in the correlation volume, that may be explained by a crossover between the rate of decrease with $T$ in the elastic an pinning energies at $T_1(H)$. A decrease of $V_c$ with increasing $B$ is also obtained. Such a crossover appears in both low and HTS superconductors near the order-disorder transition, where the decrease of $V_c$ with increasing $T$ and or $B$ is generally accompanied by an important grown of the number of dislocations, and the appearance of metastabilities and history effects. We remark that in the present case no history effects are observed near this crossover. Moreover, in that case, a crossover to a plastic creep regime should be observed. The absece of history effects and the validity of an unique critical exponent $\mu \sim 1$ in all the range is consistent with the absence of dislocations.

## VI. CONCLUSIONS

By means of a new scaling procedure, we have shown that the current relaxation in $MgB_2$ is well described by a thermal activated motion across activation energies $U(J)$ that diverge when $J$ goes to 0, characteristic of a glass response. Due to the intermediate current range $(J_c - J) \lesssim J_c$, the explicit dependence of $J_c(T)$ must be taken into account in the analysis. We have come after the description of the observed experimental behavior with an unique critical exponent $\mu \sim 1$, characteristic of collective creep in the intermediate vortex bundle regime. The estimated correlation volumes are in agreement with this picture.

In summary, we have presented a consistent description of vortex creep in $MgB_2$ system. We believe that the procedure may be useful in a brother variety of materials.

---

**Acknowledgments**

This work was supported by CONICET, ANPCyT (PICT 01250-2006), UNCuyo, and UBACyT x142/x166.

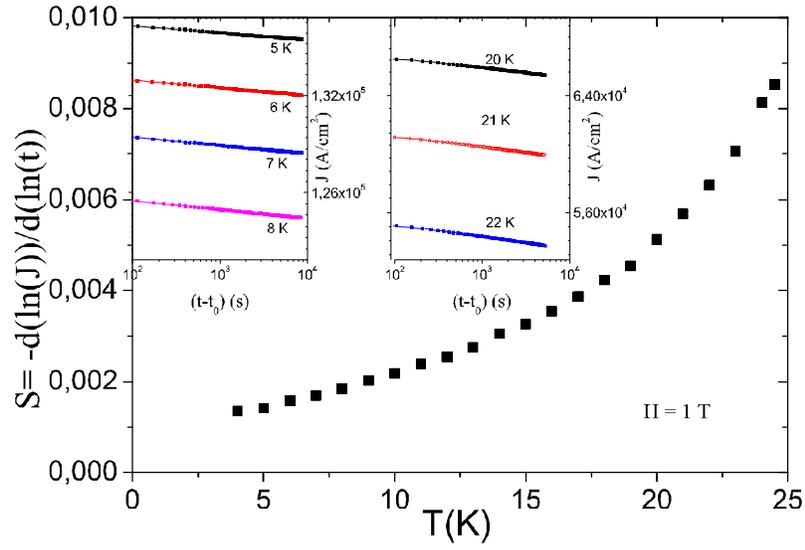

FIG. 1: Relaxation rate $S$ as a function of temperature at $H = 1T$. Insets: current density as a function of elapsed time $J(t)$ shown in double logarithmic scale, at several temperatures spaced in $\Delta T = 1K$ at low (inset a) and high (inset b).

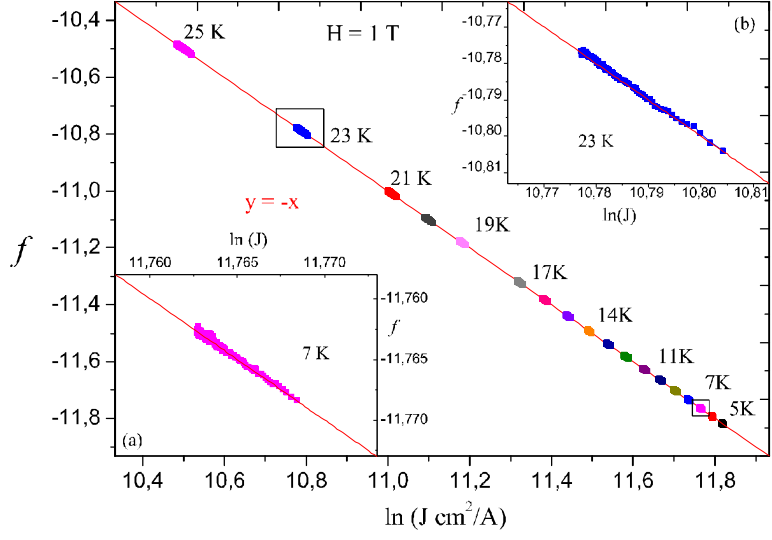

FIG. 2: Scaling function $f = \ln\left[-T\ln\left|\frac{dJ}{dt}\right| + E(T)\right] + D(T)$ obtained for relaxation data taken at $H = 1T$ with an exponent $\mu = 1.00 \pm 0.03$. The dimensionless $f$ was calculated expressing $J$ in $A/cm^2$, $t$ in $s$ and $T$ in $K$. Insets: zooms to show examples of the good fit obtained at 7 $K$(inset a) and 23 $K$ (inset b) .

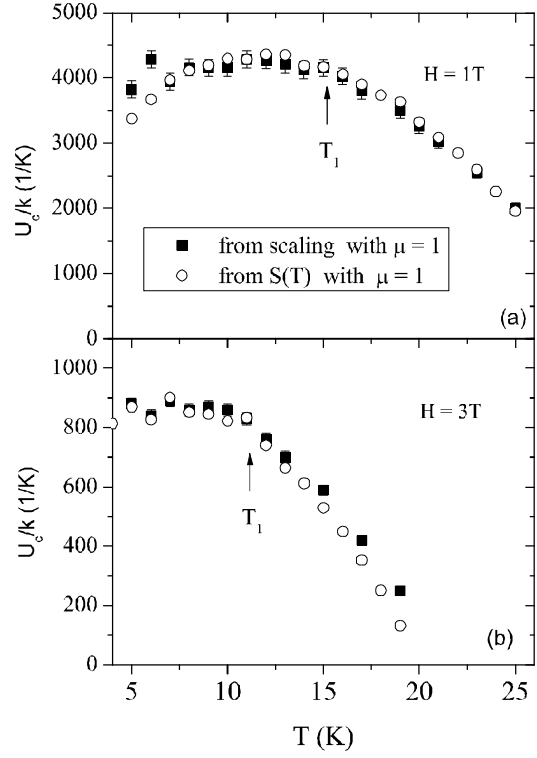

FIG. 3: Pinning energies $U_c(T)/k$ obtained from the parameters $E(T)$ (black symbols) and from the experimental $S(T)$ (hollow symbols) at $H = 1T$ (a) and $H = 3T$ (b).

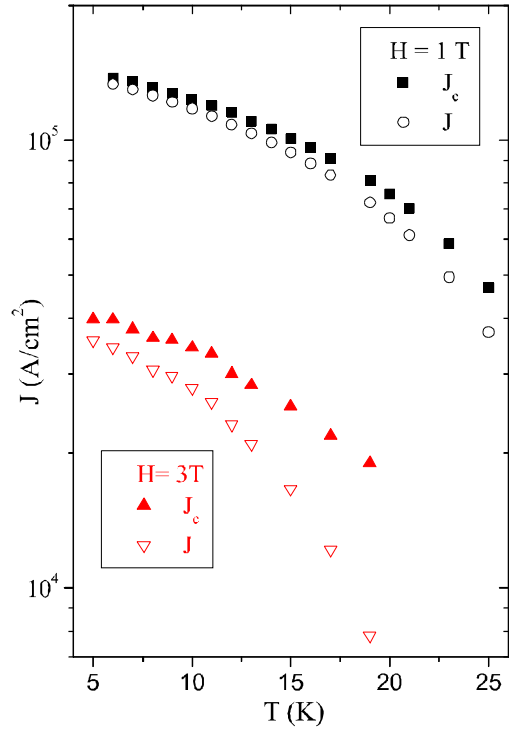

FIG. 4: Measured current density $J$ at the beginning of the relaxation (i.e."measurable" critical current) compared with the "true"$J_c$ calculated from the scaling function, at $H = 1T$ and $3T$. The condition of an intermediate creep, with $(J_c - J) \lesssim J_c$ holds in all the range.

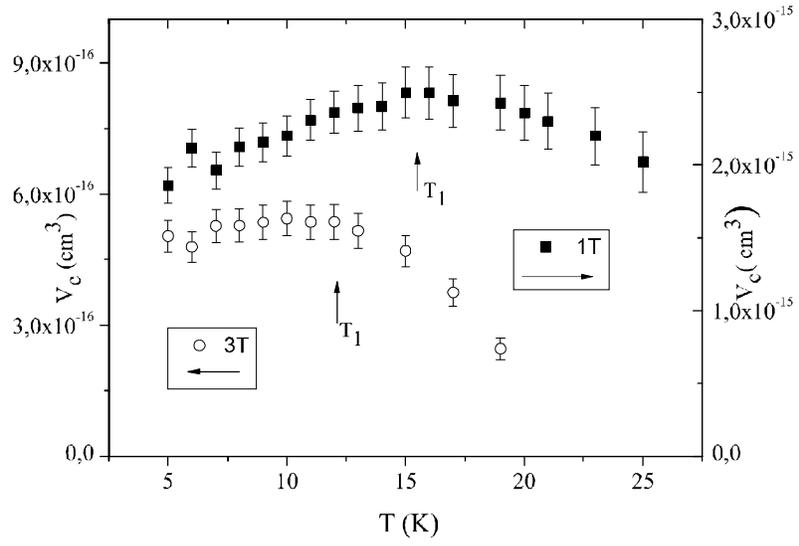

FIG. 5: Collective correlation volume $V_c$ as a function of temperature for $H = 1T$ and $3T$.